# Dislocation Engineering: A New Key to Enhancing Ceramic Performances

**Haoxuan Wang, first author**

School of Aerospace Engineering, Xi'an Jiaotong University,

710049, China

whx3122106094@stu.xjtu.edu.cn

**Yifan Wang, second author**

School of Aerospace Engineering, Xi'an Jiaotong University,

710049, China

wyf994490319@163.com

**Xu Liang[1], third author**

School of Aerospace Engineering, Xi'an Jiaotong University,

710049, China

xliang226@xjtu.edu.cn

**Wenshan Yu, fourth author**

School of Aerospace Engineering, Xi'an Jiaotong University,

710049, China

wenshan@mail.xjtu.edu.cn

**Xufei Fang[2], fifth author**

Institute for Applied Materials, Karlsruhe Institute of Technology,

76131, Germany

xufei.fang@kit.edu

**Shengping Shen, sixth author**

School of Aerospace Engineering, Xi'an Jiaotong University,

710049, China

sshen@mail.xjtu.edu.cn

[1]Corresponding author,

[2]Corresponding author,

## Abstract

Dislocations are line defects in crystalline solids and often exert a significant influence on the mechanical properties of metals. Recently, there has been a growing interest in using dislocations in ceramics to enhance materials' performance. However, dislocation engineering has frequently been deemed uncommon in ceramics owing to the brittle nature of ceramics. Contradicting this conventional view, various approaches have been used to introduce dislocations into ceramic materials without crack formation, thereby paving the way for controlled ceramics performance. However, the influence of dislocations on functional properties is equally complicated owing to the intricate structure of ceramic materials. Furthermore, despite numerous experiments and simulations investigating dislocation-controlled properties in ceramics, comprehensive reviews summarizing the effects of dislocations on ceramics are still lacking. This review focuses on some representative dislocation-controlled properties of ceramic materials, including mechanical and some key functional properties, such as transport, ferroelectricity, thermal conductivity, and superconducting properties. A brief integration of dislocations in ceramic is anticipated to offer new insights for the advancement of dislocation engineering across various disciplines.

## 1. Introduction

The performance of materials is collectively determined by their elemental composition and internal structure. However, materials inherently contain various defects including point defects [1], line defects [2], planar defects [3], and volumetric defects [4], which directly alter the atomic arrangement and significantly impact material properties. In this review, we primarily focused on the recent research of influence of line defects, specifically dislocations, on ceramics performances.

In the past, the significance of dislocations in ceramics has often been overlooked, with research predominantly concentrating on metallic materials. This can be attributed to two main factors: firstly, dislocations in ceramics typically cannot move easily compared to their counterparts in metals; secondly, the quantity of dislocations in conventionally sintered ceramics is much lower than in metals and often neglected. Moreover, dislocations have traditionally been suppressed in ceramics, as their formation and accumulation can serve as stress concentrators, potentially initiating crack nucleation and propagation within the microstructure [5].

Nowadays, ongoing research has revealed that strategically manipulating dislocations can enhance certain properties of ceramics. For example, the mechanical behavior of ceramics is profoundly affected by dislocations, which is demonstrated by their inherently high stiffness and hardness, along with their limited plasticity at room temperature. Ceramics tend to fracture much more easily under external forces, in contrast to metals, revealing extremely low toughness and high brittleness [6]. Surprisingly, it is found that effective control and management of dislocations are crucial in ceramic engineering to improve strength and toughness [7-13]. Moreover, through extensive research, it has been discovered that the strain gradient and electric field inherent at dislocation sites can induce numerous remarkable phenomena and functional applications that are challenging to achieve in pristine ceramics. For example, dislocations, are recognized for their potential to tune conductivity [14, 15], piezoelectricity [16, 17], ferroelectricity [18], flexoelectricity [19], thermal conductivity [20], superconductivity [18], and even catalytic activity [21] in ceramic materials. Therefore, the research focus on dislocations has gradually shifted from metals to other materials [22-35], especially on ceramics [7, 36-41]. Another significant aspect is that, in recent years, with the introduction of various dislocation generation methods [42-53], the creation of a large number of controllable dislocations in ceramics in a rational and orderly manner without crack formation can be achieved. However, despite the extensive experimental results have been found that various material properties can be enhanced by dislocations in functional ceramics, the underlying mechanisms for how dislocations affect the properties of ceramic materials still remain unclear.

The impact of dislocations varies with material structure and composition. Dislocation in metals was first studied, where dislocations glide easily due to dense lattices and strong metallic bonds, enhancing plasticity. In contrast, ceramics—with their covalent bonding—restrict dislocation motion, reducing plasticity and increasing brittleness. This is because the electrostatic constraints in ceramics limit the available slip systems, causing dislocations to pile up at grain boundaries and eventually initiate cracks [2]. Therefore, when studying the effects of dislocations on ceramics, one cannot directly extrapolate findings from research conducted on metallic materials. Moreover, similar conclusions may vary significantly across different ceramics. These complexities present considerable challenges in investigating the influence of dislocations on ceramic properties. Therefore, it is useful to provide an overview of the results obtained in the past years regarding the manipulation of material properties through dislocations. In this review, we aim to provide an overview of current studies on the effect of dislocation on material properties and discuss the underlying mechanisms. The content of this article is mainly focused on ceramics, and some related single crystal and thin film will also be covered.

## 2. Dislocation tuned mechanical properties

### 2.1 A brief review of dislocations

To begin with, a clear understanding of the classical dislocations is a prerequisite for grasping its effects. Dislocations are line defects in crystalline materials that form the boundary between regions of the crystal lattice where slip (shear displacement) has occurred and regions where it has not. Geometrically, a dislocation can be visualized as the line separating the slipped and un-slipped atomic planes, with the lattice distortion concentrated near this boundary. The two primary types of dislocations are edge dislocations and screw dislocations, as shown in Fig. 1(a) and (b). Mixed-type of dislocations are frequently observed in real crystals as in Fig. 1(c), possessing characteristics of both edge and screw dislocations [54].

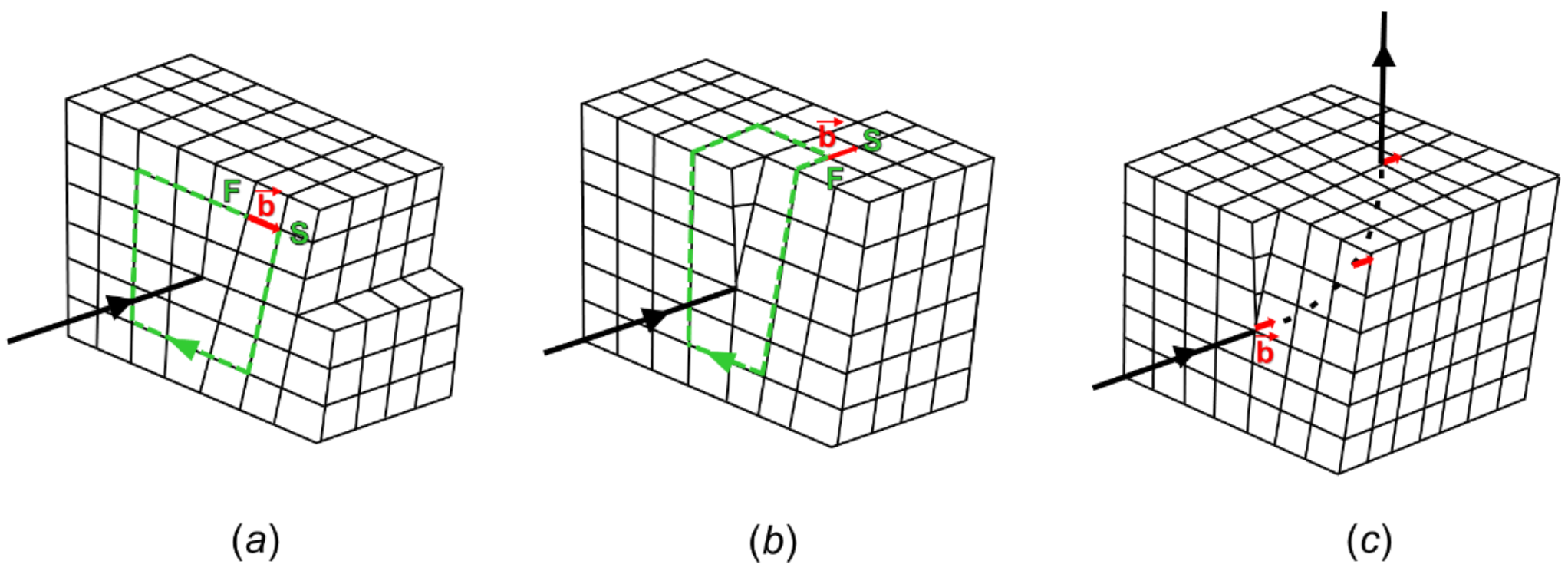

**Fig. 1** The elemental models of (a) edge, (b) screw and (c) mixed dislocations in a simple cubic structure. Black arrows indicate dislocation line. Burgers circuit is marked in green (S-start and F-finish positions of atom-to-atom counting). The Burgers vector **b** is shown in red.

The study of dislocations can be traced back to the previous century, with transmission electron microscopy (TEM) being the most widely employed technique for their observation [55]. Meanwhile, other experimental techniques including surface methods [56], decoration methods [57], X-ray diffraction (XRD) [58], field ion microscopy [59], and atomic probe tomography [60] have also been developed to unravel the mysteries of dislocations. However, owing to the atomic-scale nature of dislocations, many of their properties are challenging to observe and analyze experimentally. Therefore, dislocation engineering did not receive much attention at that time. Recent years, with the rapid advancement of computational simulation techniques, researchers have conducted more in-depth studies on dislocations through simulation and analysis. In the context of the simulations of dislocations, the computational tools in this framework, in order from smallest to largest scale, are quantum mechanics, classical molecular dynamics (MD), discrete dislocation dynamics (DDD), continuum dislocation dynamics (CDD), and crystal plasticity (CP) models [61-64]. The development of science and technology has provided a better platform for researchers to study dislocation engineering more effectively.

### 2.2 The influence of dislocations on the mechanical properties of ceramics

Dislocations play a central role in governing the mechanical behavior of crystalline materials. Their nucleation and movement provide the primary mechanism for plastic deformation, significantly lowering the required shear stress compared to the theoretical value in perfect crystals. Moreover, the interaction and entanglement of dislocations will lead to strain hardening, whereas their limited mobility in ceramics and covalently bonded materials contributes to brittleness [65]. In cyclic or high-temperature conditions, dislocations also control fatigue and creep behaviors, respectively, making their presence a crucial factor across multiple deformation regimes [66, 67].

To explore the impact of dislocation on the mechanical performance of ceramics, it is essential to comprehend their modes of motion [6, 68]. Slip is the most prevalent mode of dislocation movement in materials, referring to the atomic layer sliding along crystallographic planes within the crystal lattice, resulting in the displacement of dislocation lines. Dislocation slip involves the gliding of one atomic plane over another, typically along the direction of maximum atomic packing (densest planes). The slip direction depends on lattice orientation, and multiple crystallographic slip systems may operate in a crystal. However, it should be noted that the slip mechanism in ceramics is typically much more complex, as they lack the specific guidelines of fixed close-packed planes and directions that metals possess. This is because in metals, all atoms carry a positive charge and the electrons can freely move, allowing slip to occur along the easiest slip direction, which is the direction of highest packing density. However, in ceramics, atomic structure is typically composed of ions or covalent bonds, making it impossible to bring together ions with like charges, which limits the slip planes and directions [69]. Additionally, other factors such as width of the dislocation and Burgers vector can also influence dislocation motion. The Peierls -Nabarro equation is used to describe the resistance to motion by a screw dislocation.

$$\sigma_{\mathrm{P}} = \frac{2\mu}{1-v}\left(-\frac{2\pi d}{b(1-v)}\right) \quad (1)$$

where $\sigma_{\mathrm{P}}$ is called Peierls-Nabarro stress, which is the driving force of dislocation. In particular, $\mu$ is the shear modulus, $v$ and $d$ represents the Poisson’s ratio and the spacing between slip planes, respectively. And $b$ is the magnitude of the Burgers vector.

In ceramics, the movement of dislocations is typically hindered due to their narrow widths and high resistance. Further calculations revealed that the resistance to dislocation motion is heavily influenced by the interatomic forces. However, precise calculations of this resistance require a better understanding of the laws governing interatomic forces [70, 71]. The slip systems of various ceramics need to be considered individually, taking single-crystal magnesium oxide (MgO) and strontium titanate ($SrTiO_3$) as examples. MgO features a simple rock-salt-type structure, also known as a sodium chloride-type (NaCl) structure, belonging to the cubic crystal system with a face-centered cubic lattice and space group of Fm3m. Owing to its simple structure, observations and studies conducted half a century ago using in-situ atomic force microscopy (AFM) and transmission electron microscopy have established that the slip planes in MgO are {110} or {100} (at high temperature), with the slip direction being <110> [72-74], which was also verified by computational materials sciences [75], the slip system of MgO is shown in Fig. 2. Nevertheless, for ceramic materials possessing more complex structures, such as perovskites, the intricacies of their slip systems prompt widespread debate. Taking single-crystal $SrTiO_3$ as an example, Yang et al. suggests that the slip system at room temperature are <110> {001} and <110> {110} [76]. However, their proposed slip system <110> {001} lacks additional experimental evidence, as Matsunaga et al. and Kondo et al. only observed <110> $\{1\bar{1}0\}$ in their experiments at room temperature without detecting <110> {001} [77, 78], and <110> $\{1\bar{1}0\}$ slip system is only active at high temperature [79]. Additionally, since certain slip systems cannot be experimentally observed at room temperature, such as <100> {001}, researchers have turned to molecular dynamics to study these slip systems [80]. Different materials have distinct slip planes and directions, and due to experimental limitations for certain slip systems, simulations have become a crucial tool for investigating slip systems [81-83].

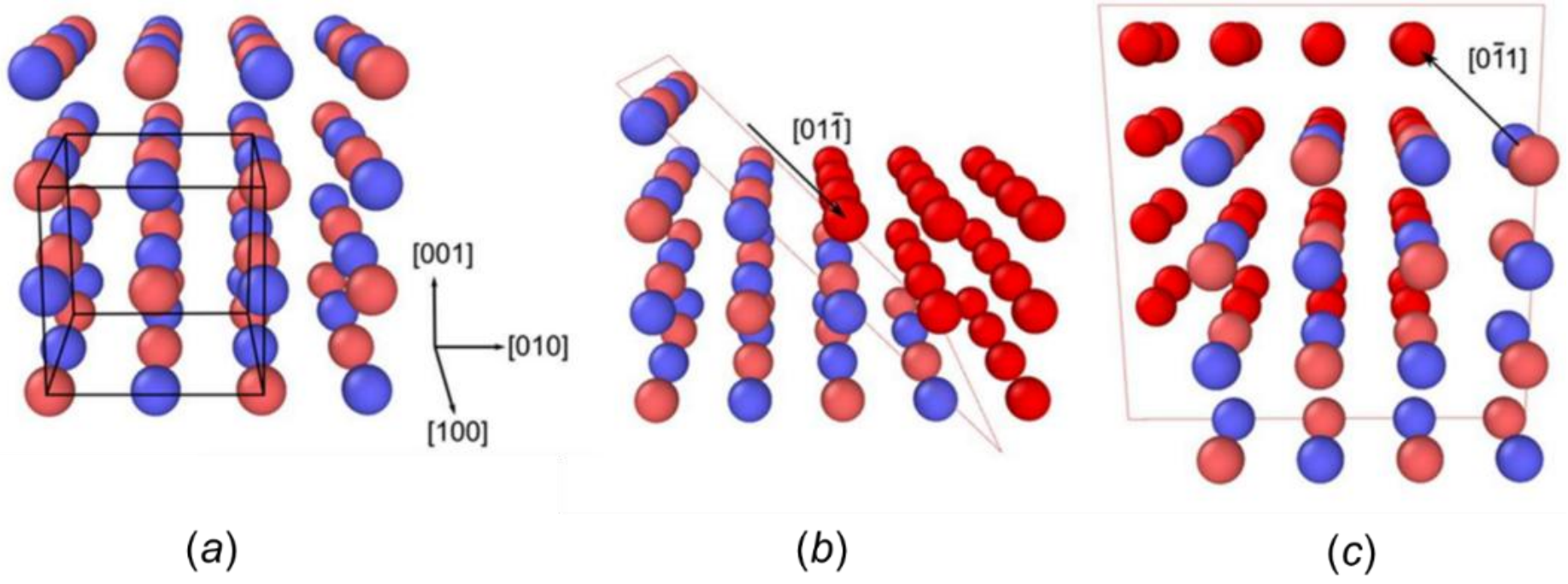

**Fig. 2** (a) Perspective view of the MgO crystalline structure. The unit cell is drawn with black lines and the crystal is oriented along the cubic orientation. (b, c) Illustrations of the shearing process for the (b) <110> {110} and (c) <110> {100} slip systems. The same orientation is used and the atoms displaced during the shearing process are labeled in red. Black arrows refer to typical Burgers directions [74].

It is imperative to acknowledge that due to the presence of ionic or covalent bonds in the atomic structure of ceramic materials, the interatomic bonding is inherently stronger. Consequently, slip predominantly occurs under conditions of elevated temperature and high stress. Within these circumstances, dislocations are able to traverse along the slip planes of the crystal, forming slip bands that result in visible plastic deformation. Conversely, at room temperature and under low stress conditions, ceramic materials tend to fracture rather than undergo plastic deformation [84, 85]. This arises from the fact that the extent of plastic deformation in a material relies on the density of dislocations, their Burgers vectors, and the distance over which dislocation movement occurs. The plastic strain is

$$\gamma = \varsigma b \lambda, \qquad (2)$$

where $\varsigma$ is the mobile dislocation density, $b$ and $\lambda$ is the Burgers vectors and the average slip distance, respectively. At room temperature, the distance over which dislocations in ceramics can move is limited due to their inherent short-range motion capability. Additionally, the dislocation density in ceramics is typically low (e.g., $10^9$ $m^{-2}$ for ceramic [47] and $10^{12}$~$10^{14}$ $m^{-2}$ for metal [86]). Consequently, when subjected to external forces, ceramics typically undergo fracture rather than plastic deformation.

This raises an important question: is it possible to manipulate the mechanical properties of ceramics without introducing cracks? Recent advances suggest that it is indeed achievable through dislocation engineering.

## 2.3 The approach to tune ceramic mechanical properties

### *2.3.1 Processing*

Dislocations in ceramic materials can be generated through several distinct mechanisms, depending on both intrinsic and extrinsic factors. The dislocations generated in $SrTiO_3$ under different processing conditions are shown in Fig. 3.

One of the primary sources of dislocations in ceramics is the intrinsic defect formation during crystal growth. For instance, adjusting the doping ratio during the processing of ceramics can effectively control the overall material's defect formation. Nakamura et al. achieved significant improvement in the plasticity of $SrTiO_3$ ceramics by altering the Sr/Ti doping ratio, leading to high plasticity with strains exceeding 10% at room temperature [87]. Subsequent studies have revealed that by further optimizing the doping ratio, a higher proportion can result in the generation of a significant number of oxygen vacancies [88]. These vacancies facilitate easier nucleation of dislocations, thereby reducing the maximum shear stress and altering the material's plasticity.

Emerging sintering techniques also offer opportunities to regulate the mechanical behavior of ceramics by generating a higher dislocation density compared to conventional sintering methods. Notable examples include cold sintering (CS) of $(K_{0.5}Na_{0.5})NbO_3$ [89, 90], flash sintering(FS) of yttria-stabilized zirconia [91] and $TiO_2$ [9], oscillatory pressure sintered (OPS) of WC-$ZrO_2$-$Al_2O_3$ [92], as well as spark plasma sintering (SPS) of $ZrB_2$ [93], and alumina–SiC [94]. However, despite the variety of available sintering techniques, the resulting polycrystalline samples often exhibit defective microstructures containing pores and microcracks. These microstructural instabilities fundamentally limit sintering-based approaches for precise dislocation engineering.

A bi-crystal is a model system consisting of two single crystals joined along a well-defined grain boundary with a specific misorientation. The key advantages of the bi-crystal structure are its ability to induce dislocations in a controlled manner. By carefully selecting the misorientation angle and the boundary plane, dislocations with specific character (edge, screw, or mixed) and Burgers vector can be introduced [2]. This makes the bi-crystal configuration particularly suitable for investigating the isolated effects of single dislocations on material properties and for exploring dislocation–grain boundary interactions at the atomic scale [42, 95-97]. For example, Choi et al. used $SrTiO_3$ bi-crystals with controlled tilt angles to introduce periodic edge dislocations, showing that strain between dislocation cores promotes oxygen vacancy formation and affects electrical behavior [95]. Similarly, Feng et al., investigated YSZ bi-crystals and found that dislocation-induced electrostatic fields can repel oxygen vacancies from grain boundary cores, revealing strong dislocation-defect interactions at the atomic scale [96]. Recently, researchers have even "borrowed" dislocations from metals to enhance ceramic ductility. By designing coherent $La_2O_3$-Mo interfaces, dislocations can transfer from the metal to the ceramic, overcoming ceramics' intrinsic brittleness and achieving metal-like plasticity [98].

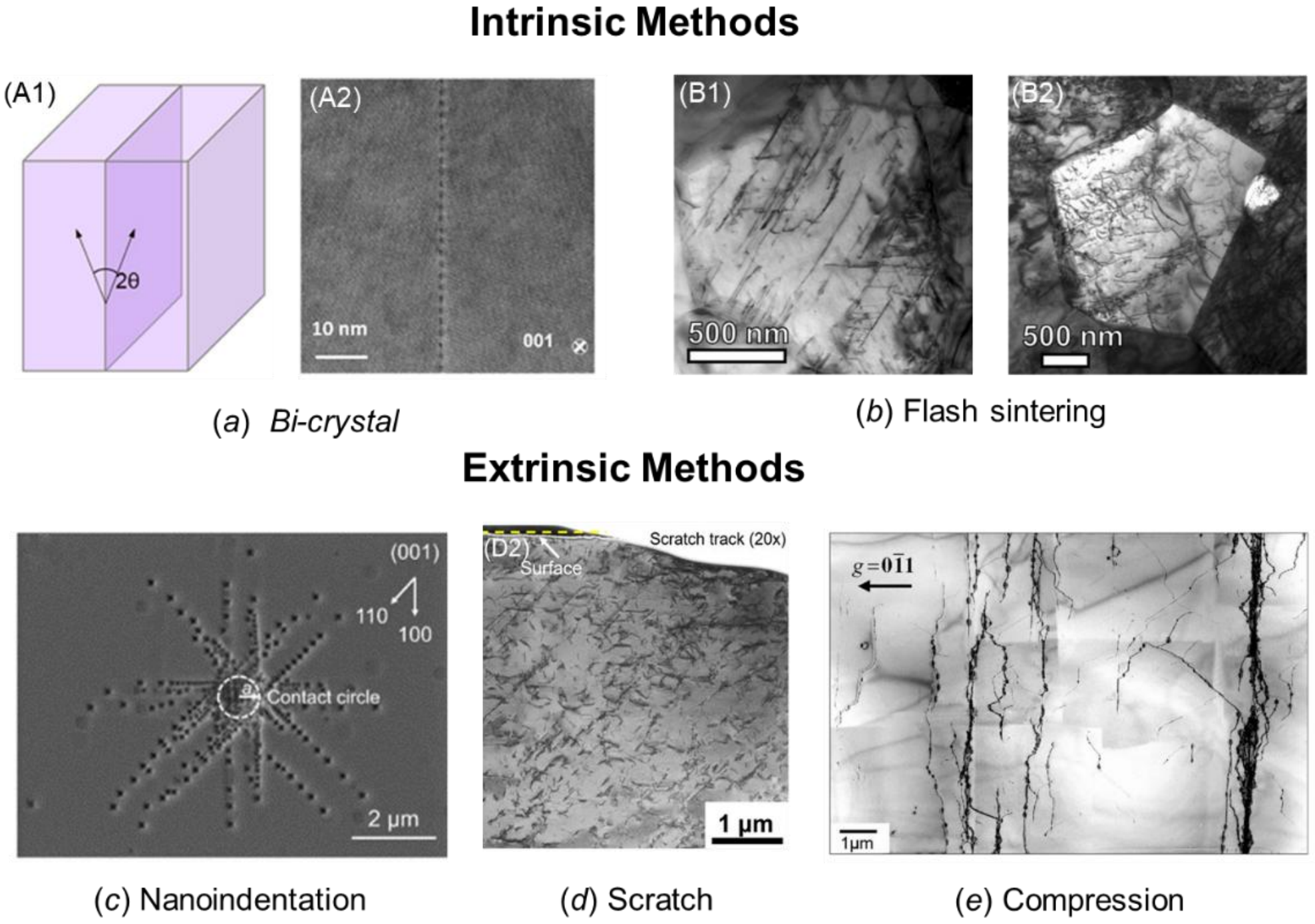

**Fig. 3** Methods for introducing dislocations into $SrTiO_3$. (a) Bi-crystal with a 10° tilt [99]. (b) Flash sintering [50]. (c) Nanoindentation test at room temperature [100]. (d) Scratch test (20 cycles) at room temperature on [49]. (e) Compression to 1% plastic strain at 815K [11].

### *2.3.2 Mechanical loading*

When a ceramic material is subjected to mechanical loading—such as indentation, bending, or high-pressure compression—localized stresses can reach levels sufficient to nucleate dislocations, especially at stress concentrators like grain boundaries, surface flaws, or pre-existing microcracks [43, 78, 79, 100, 101]. Single crystal ceramics are frequently chosen for these experiments because their simple structure allows for easy observation and analysis of slip systems and dislocations when subjected to external forces. Furthermore, some single crystals exhibit remarkable plastic deformation capabilities even at room temperature [74, 102, 103]. For instance, at room temperature, $KNbO_3$ can undergo plastic deformation of up to 5% with shear stresses of 26-30 MPa along the $<1\bar{1}0>\{110\}$ slip system [104], slightly lower than that of $SrTiO_3$ (64 MPa) under the same slip system [105]. In 2021, Porz et al. reported that by increasing the dislocation density to approximately $10^{15}$ $m^{-2}$ in micro-regions of single-crystal $SrTiO_3$, the critical stress for its plastic deformation decreased by at least 10 times [13].

High-temperature compression was used to create dislocations in ceramic materials in earlier studies [11, 43]. However, the method often suffered from low success rates due to the brittle nature of ceramics and the difficulties in achieving controlled plastic deformation at elevated temperatures. The process required extremely precise control over both temperature and mechanical load, as well as specialized equipment capable of maintaining stable high-temperature and high-pressure environments. Surprisingly, a new method emerged that enabled the creation of dislocations at room temperature in a more convenient manner — the cyclic indentation experiments performed by Okafor et al. on single-crystal $SrTiO_3$ using a Vickers hardness tester [47]. They found that, while controlling the applied load and number of cycles, the dislocation density in the indentation area reached saturation and observed that the material's hardness increased with increasing dislocation density as shown in Fig. 4(a) and 4(b). This finding was consistent with previous experimental conclusions and was attributed to the cumulative effect of dislocation stress fields [106]. Salem et al. later employed a similar approach [107], while also controlling the temperature, and discovered that increasing the dislocation density simultaneously enhanced the hardness and fracture toughness of single-crystal $SrTiO_3$, as shown in Fig. 4(c) and 4(d), despite their generally considered contradictory relationship [108]. Moreover, this method exhibits universality. In 2023, Preuß et al. conducted experiments using the same method on $KNbO_3$ and found that its fracture toughness increased by 2.8 times [7]. Recently, another new mechanical method, named "dislocation seeding", was proposed to improve the room-temperature plasticity of ceramics [49, 52, 109]. By micropillar compression tests, the density of dislocations has reached $10^{14}$ $m^{-2}$ and the brittle STO showed fascinating plasticity [52].

All these experiments provide feasible strategies for controlling the mechanical behavior of ceramics through dislocation manipulation.

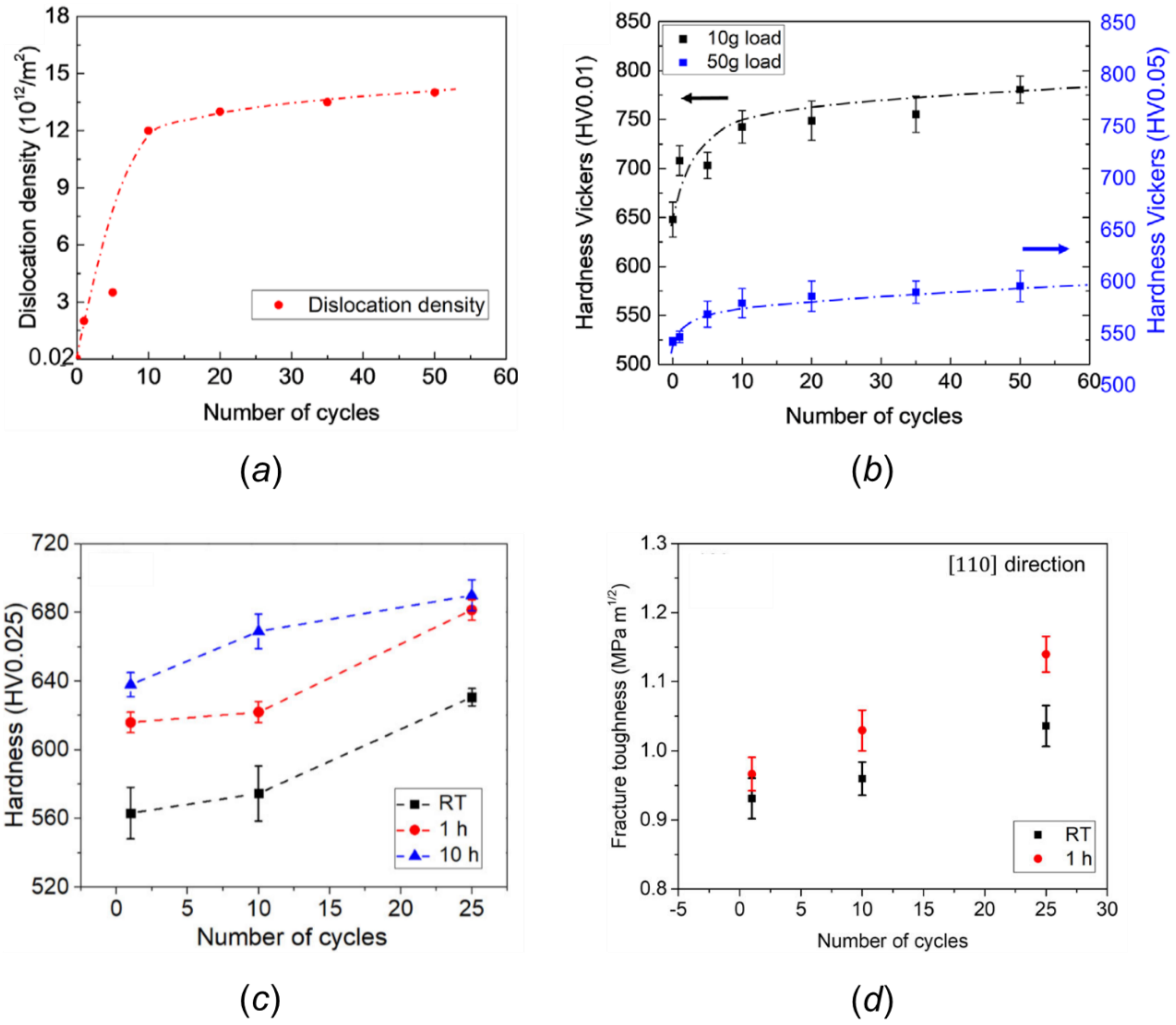

**Fig. 4** The mechanical properties of single crystal STO controlled by cyclic indentation experiments. **(**a) Plot of dislocation density in the center of the plastic zone as a function of the number of cycles. (b) Vickers micro-hardness for two different loads (10 and 50 g) in the plastic zones after a different number of cycles [47]. (c) Hardness and (d) Fracture toughness test for STO between room temperature (RT) and annealing times of 1h or 10h [107].

### *2.3.3 Controlling the motion of dislocations*

In addition to generating dislocations, it is equally important to control their mobility to fully exploit their role in modifying mechanical properties. Dislocation motion can be influenced by several external factors, with temperature being the most fundamental and accessible parameter. At elevated temperature, the interatomic forces encountered by dislocations weaken, thus rendering their movement easier. Consequently, ceramics will undergo a transition from brittle to ductile behavior (referred to brittle-to-ductile transition or BDT) [110-112]. The occurrence of BDT is influenced by numerous factors, including the inherent properties of the material, experimental conditions, and loading rate. It is also noteworthy that certain materials can exhibit plasticity when the temperature is lowered, such as single crystal $SrTiO_3$. Compression of single crystal $SrTiO_3$ led to plastic deformation within two temperature ranges: from 78 K to 1050 K and from 1500 K to 1800 K. The material was brittle at temperature falling between these two ranges (1050 K to 1500 K), as depicted in Fig. 5. It was believed that edge dislocations in $SrTiO_3$ undergo climb at high temperatures, while screw dislocations became sessile at low temperatures due to the temperature dependence of dislocation core structures [11, 113, 114].

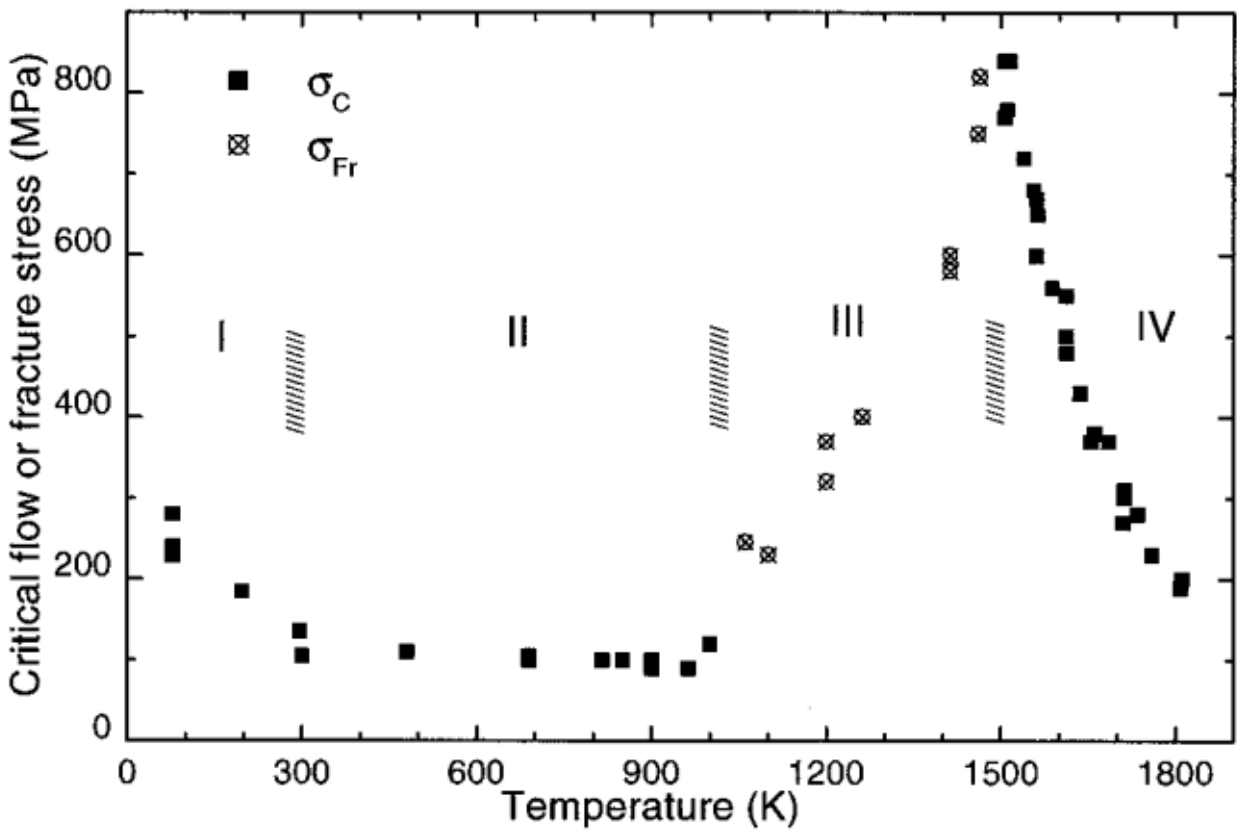

**Fig. 5** Critical flow stress ($\sigma_C$) and fracture stress ($\sigma_{Fr}$), as a function of temperature [113].

Another method involves manipulating the microstructure of the material to control the motion of dislocations, thereby enhancing the mechanical properties of the material. In polycrystalline materials, the motion of dislocations is significantly influenced by the size of the grains and the number of grain boundaries. Although it has certain limitations, the Hall-Petch relationship represents a commonly used equation for studying the impact of grain boundaries on the mechanical properties of materials [115, 116]. The yield stress, denoted as $\sigma_y$, can be expressed as

$$\sigma_y = \sigma_0 + k_1 D_{\mathrm{GB}}^{-1/2} \tag{3}$$

where $\sigma_0$ and $k_1$ are constants, and $D_{GB}$ is the average grain size. As the grain size of a material decreases, the number of grain boundaries increases. According to the equation, this results in higher yield strength. The motion of dislocations is impeded at grain boundaries, causing them to become trapped or pile up at these boundaries, as shown in Fig. 6. As a result, dislocation pile-ups occur, enhancing the strength and hardness of the material [117]. Additionally, although grain boundaries are considered barriers to dislocation motion, in ceramic materials such as $SrTiO_3$, they may not effectively impede crack propagation. In fact, grain boundaries can serve as preferential paths for crack growth due to their structural features and the tendency for intergranular fracture. In reality, this relationship is quite complex, and further reading on the topic can provide readers with a better understanding [29, 101, 118, 119].

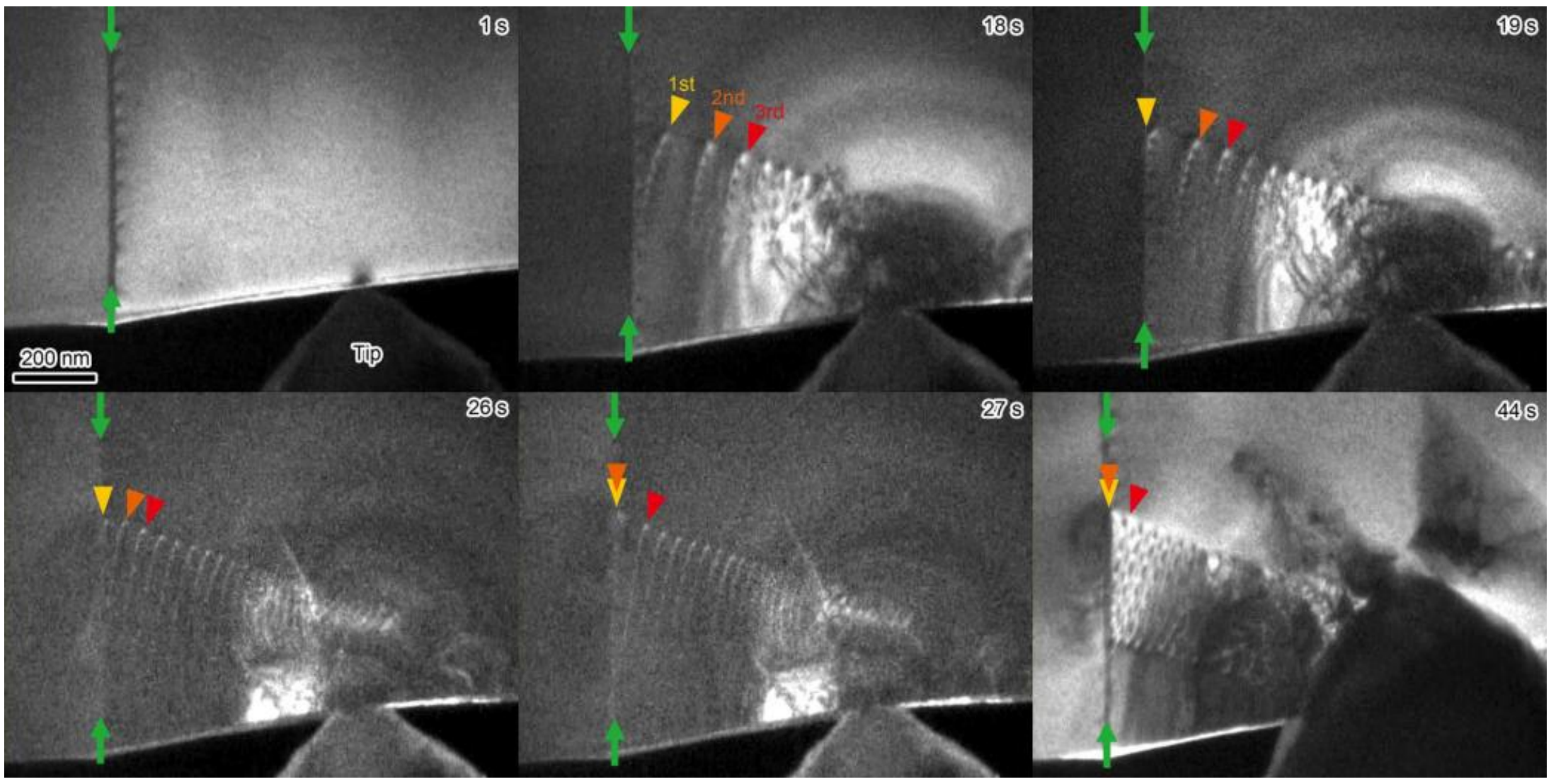

**Fig. 6** Sequential dark-field TEM images from the nanoindentation experiment on the Σ5 grain boundary. The green arrows indicate the line contrasts corresponding to the Σ5 grain boundary, while the triangles mark the positions of the first three lattice dislocations. The indenter tip moved gradually from 0 to 42s, with specimen fracture occurring at 43s. Dislocation motion was strongly impeded by the grain boundary, resulting in pileup, with the first and second dislocations and the lower part of the third dislocation trapped on the grain boundary plane after the external stress was removed [101].

### *2.3.4 Illumination*

The other means of controlling the mechanical properties of materials is through illumination. For example, Oshima et al. discovered that in a specific inorganic semiconductor material [10], ZnS, plasticity will significantly increase if under dark conditions, with $\varepsilon_t$ reaching up to 45%, as shown is Fig. 7. This was attributed to the influence of dislocations within ZnS on the optical bandgap of the entire crystal, thereby altering its plasticity.

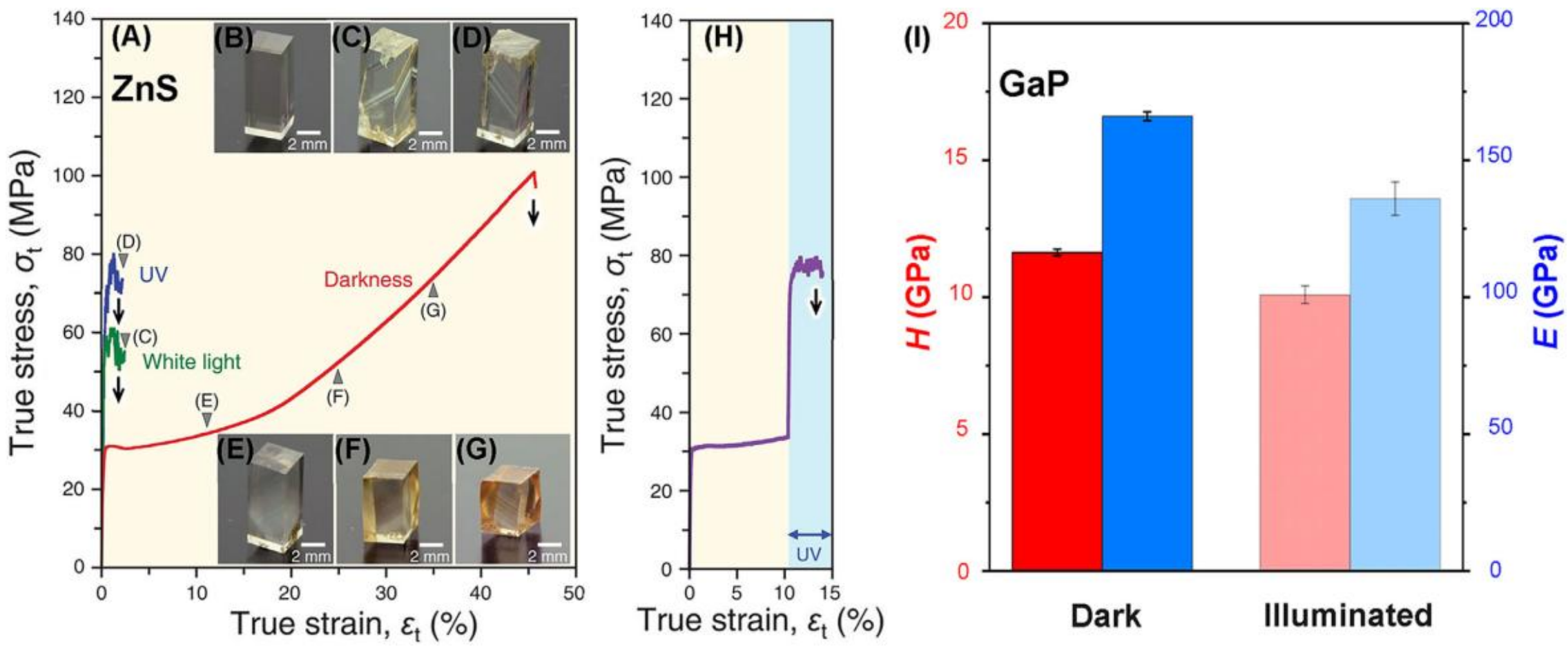

**Fig. 7** Characterizations of plastic deformation. (A) Stress-strain curves of ZnS single crystals under white or UV light (365 nm) or in complete darkness. (B) An undeformed specimen. (C and D) The specimens deformed under (C) white light-emitting diode (LED) light and (D) UV LED light (365 nm). (E to G) The specimens deformed up to (E) $\varepsilon_t$ = 11%, (F) $\varepsilon_t$ = 25%, and (G) $\varepsilon_t$ = 35% in complete darkness [10].

The study of the influence of dislocations on the mechanical behavior of ceramics has been an active topic for many years, with a wealth of experimental results confirming that dislocations can enhance the mechanical properties of ceramics. However, we have found that some of the existing mature approaches lack attempts at application in other directions. For instance, researchers have discovered that altering doping ratios can affect the quantity of dislocations, thereby changing toughness, but there is limited research on the changes in other properties.

## 3. The effect of dislocation on functionalities

From around the year 2000, research into the functional modulation of ceramics through dislocations began to gain traction. Previously, ceramics were typically sintered at high temperatures, resulting in dense structures with minimal dislocations. However, as mentioned earlier, advancements in techniques have made it feasible to introduce dislocations into ceramic materials. Consequently, manipulating material functionality through dislocations has emerged as a current research focus.

In the following sections, we will explore how dislocations influence the functionality of ceramics in four areas: transport properties, thermal conductivity, ferroelectricity and superconductivity.

### *3.1 Transport properties*

As widely acknowledged, dislocation lines can serve as rapid pathways for atomic diffusion in metals, a fundamental phenomenon referred to as pipe diffusion. This phenomenon has been substantiated through a combination of theoretical and experimental approaches in various metal [120-122] and alloy materials [123, 124], highlighting its crucial role in understanding and manipulating diffusion mechanisms in these materials. Therefore, the question arises as to whether pipe diffusion phenomenon also occur in ceramics and whether basic insulating ceramic materials can be rendered conductive through dislocations. However, research on transport properties in ceramics is still in its preliminary stages, and there are many questions that need to be addressed. For instance, what is the interaction between dislocations and point defects? Since transport phenomena depend on the nature of point defects, is the conductivity of the material derived from electrons or ions? Another question is whether the structure of ceramic materials can influence the transport phenomena of dislocations, and whether different chemical compositions of ceramics with the same structure can also have an impact. Additionally, the manner in which dislocations are generated, as well as the orientation of dislocations, may also affect the transport phenomena in the material.

The initial material explored for ceramic ion transport was yttria-stabilized zirconia (YSZ). This ceramic material, achieved by incorporating suitable amounts of yttrium oxide into zirconium oxide, exhibits elevated levels of ionic conductivity, which plays a significant role in applications such as solid oxide fuel cells (SOFCs) and oxygen sensors. The conductivity of YSZ primarily stems from the migration of oxygen vacancies. Consequently, researchers naturally wondered whether dislocations could play a similar facilitating role in the ionic conductivity of YSZ, as they do in metallic materials. Initially, Otsuka et al. conducted compression experiments on single crystal YSZ at 1300°C, resulting in a dislocation density of $8\times10^{12}$ $m^{-2}$ [15, 125]. They observed that the conductivity of the deformed sample was significantly higher than that of the undeformed sample, as shown in Fig. 8(a), attributing this enhancement to dislocations acting as a result of pipe diffusion. Their explanation posited that during plastic deformation, the migration enthalpy of oxygen vacancies decreases while the associated enthalpy increases. Consequently, oxygen vacancies surrounding dislocations might rapidly move along the dislocation lines, thereby enhancing the ionic conductivity. Furthermore, the experimental phenomenon also applied to sapphire in terms of enhancing electronic conductivity [126].

Subsequently, researchers turned their attention to studying the influence of dislocations on thin films and multilayered YSZ synthesized on different substrates. This is because when thin films are generated with materials having incomplete lattice parameter matching, interface mismatch dislocations occur in the resulting thin films. Therefore, investigating the impact of such dislocations on the performance of thin film materials is of great significance. However, there has been intense debate regarding whether these dislocations have any influence on the conductivity performance and their underlying mechanisms. For example, Sillassen et al. discovered that when epitaxial cubic YSZ is synthesized on single crystal STO and MgO substrates, a high density of misfit dislocations appears at the semicoherent interface with the MgO underlayer [127]. This results in enhanced conductivity, even reaching a superionic conductivity level (~1 $\Omega^{-1}cm^{-1}$). However, Sayle et al. through atomistic simulation found that the heterolayer $CeO_2$/YSZ system did not exhibit exceptional transport properties compared to the parent materials [128]. Li et al. investigated multilayered films of 8 mol% YSZ with $Gd_2Zr_2O_7$ (GZO) and discovered that fewer dislocations could significantly enhance the ionic conductivity [129]. Harrington et al. comprehensively analyzed lattice strain, dislocations, and microstructure's impact on material transport properties in YSZ films grown on different substrates [130]. Their findings provided no evidence to support the notion that dislocations can enhance transport rates. They argued that film texture and grain boundary density dictated the transport properties in the films. Similarly, Schichtel et al. also suggested that an increased density of dislocations can lead to an improvement in the conductivity of YSZ [131, 132]. However, they attributed this behavior to strain fields rather than pipe diffusion. Recently, Feng et al. analyzed the phenomenon of enhanced ion conduction associated with a single crystal YSZ's edge dislocation from a structural perspective [133]. They proposed that the coupling between tensile strain fields and compositional changes around the dislocation core created a faster ion conduction pathway in the vicinity of the dislocation core. Currently, there is little controversy about the significant enhancement of conductivity achieved by generating dislocations in single-crystal YSZ samples through high-temperature tensile testing. However, considerable debate remains regarding the impact of dislocations on the conductivity of YSZ films grown on various substrates and in multilayer YSZ films, as well as the mechanisms underlying their enhancement or reduction.

Another branch of research is the perovskite material $SrTiO_3$ (STO) with mixed ionic-electronic conduction [36, 134, 135]. Due to the ability to correlate experimental and simulation results, there have been numerous and convincing studies on the transport properties of STO, focusing primarily on the phenomenon of resistive switching in STO. Szot et al. found that under low oxygen partial pressure conditions at 800°C, thermal reduction of single crystal STO led to a significant increase in dislocation density [14, 136]. Consequently, the single crystal STO exhibited metallic conductivity, which decreased as the dislocation density decreased, as shown in Fig. 8(b). This finding demonstrates that dislocations can enhance the conductivity of STO, which they attributed to the dislocation providing an easier pathway for oxygen transport. Subsequently, they further investigated individual dislocations in single crystal STO and discovered the phenomenon of resistive switching. In the metal-STO-metal structure, the key to achieving resistive switching lies in the electroforming step. During this step, the application of a higher voltage or current leads to the migration of oxygen ions and the formation of oxygen vacancies in $SrTiO_3$. These oxygen vacancies then form a highly conductive filament, resulting in a significant decrease in resistance in $SrTiO_3$, thus achieving resistive switching. The formation of this filament is reversible and can be disrupted by applying a reverse voltage or current. The authors suggest that this is a localized pipe diffusion phenomenon in STO which allows for the localized modulation of oxygen content. However, this viewpoint was later disproven by Metlenko et al. through oxygen tracer diffusion studies, which did not find evidence of rapid oxygen ion diffusion along dislocations [137]. Additionally, due to the higher activation energy required for long-distance migration, it appears that the movement of oxygen ions along dislocations is hindered. These findings were subsequently confirmed by Stephan et al. and Schraknepper et al, explaining that oxygen ion transport was hindered by dislocations [138, 139]. However, what exactly is the principle behind the phenomenon of resistive switching? Marrocchelli et al. suggest that the local changes in electrical properties caused by dislocations are simply due to the easier reduction to form oxygen vacancies [140]. Despite the lower mobility of oxygen ions compared to those in the bulk, the accumulation of a large number of oxygen vacancies leads to higher diffusion coefficient. Rodenbücher et al. propose that the accumulation of electrons in space charge tubes around dislocations provides a pathway for current to flow [141]. This current, through Joule heating, increases the temperature of the dislocations and the surrounding material. Outside the space charge tubes, there are sufficient oxygen vacancies, and with the increase in temperature, their migration rate increases dramatically, leading to oxygen ion conduction and resulting in nanoscale redox reactions that generate conductive filaments. By integrating electrical and chemical tracer diffusion experiments with computational models, Adepalli et al. revealed that dislocations alter the equilibrium concentration and distribution of electronic and ionic defects [142]. The presence of overlapping electrostatic fields around dislocation cores with positive charges has been found to affect diffusion coefficients and electronic conductivity. Furthermore, Lukas Porz et al. conducted a series of comprehensive experiments on the conductivity of dislocation-engineered strontium titanate single crystals [36]. They concluded that it is a combined effect of mesoscopic structure, core, and space charge. In conclusion, it has been found that the conductivity of strontium titanate can be modulated by dislocations, and the underlying mechanism is not simply a pipe diffusion like in metals, but rather a combined effect of ionic and electronic defects. To better understand the research progress of transport properties of $SrTiO_3$ modified by dislocation, the studies conducted since the 21st century were briefly summarized, as shown in Table 1.

**Table 1.** Summary of recent work on the effect of dislocations on transport properties of $SrTiO_3$

| Year | Effect on transport | Explanation | Ref |
|---|---|---|---|
| 2002 | metallic conductivity, which decreased as the dislocation density decreased | easy diffusion paths for oxygen | [14] |
| 2006 | resistive switching | pipe diffusion | [136] |
| 2015 | vacancy accumulation at the dislocation core for the case of low vacancy concentration | the easier reduction to form oxygen vacancies | [140] |
| 2017 | the n-type conductivity increases by 50 times at oxygen pressures below $10^{-5}$ atm; the p-type conductivity decreases by 50 times at higher oxygen pressures | overlapping electrostatic fields around dislocation cores with positive charges | [142] |
| 2018 | the damaged zone constitutes an additional resistance to the transport of oxygen in and out of the sample | enclosed in space-charge tubes depleted of oxygen vacancies. | [139] |
| 2019 | n-type conductivity increases by a factor of $10^4$ in the vicinity of dislocations | filamentary conduction along dislocations under reducing conditions | [141] |
| 2020 | dependent on mesoscopic structure of defects | a combined effect of mesoscopic structure, core, and space charge | [36] |

In other materials such as $TiO_2$, it has been observed that dislocations can also modulate conductivity. Sun et al. explained this phenomenon as a confinement of spin-polarized conducting states within the dislocation core, forming spatially connected pathways for conduction [143]. Adepalli et al suggest that the observed phenomenon of dislocation-induced conductivity in $TiO_2$ can be attributed to an increase in titanium vacancy content within the dislocation core [37], the impedance change of $TiO_2$ with and without dislocations is shown in Fig. 8(c). Maras et al. proposed that dislocations can act as electron traps [144]. Muhammad et al. developed a computational model describing dislocation bundles as overlapping regions of charge [145]. Currently, efforts have been made to utilize this phenomenon for local or global control of conductivity. For example, nanoindentation can generate blade-shaped dislocations in small regions of single crystal $TiO_2$, leading to a 50% enhancement in conductivity within the dislocation-rich area [146]. Additionally, spark plasma sintering has been utilized to produce polycrystalline $TiO_2$ with abundant dislocations for conductivity modulation [147].

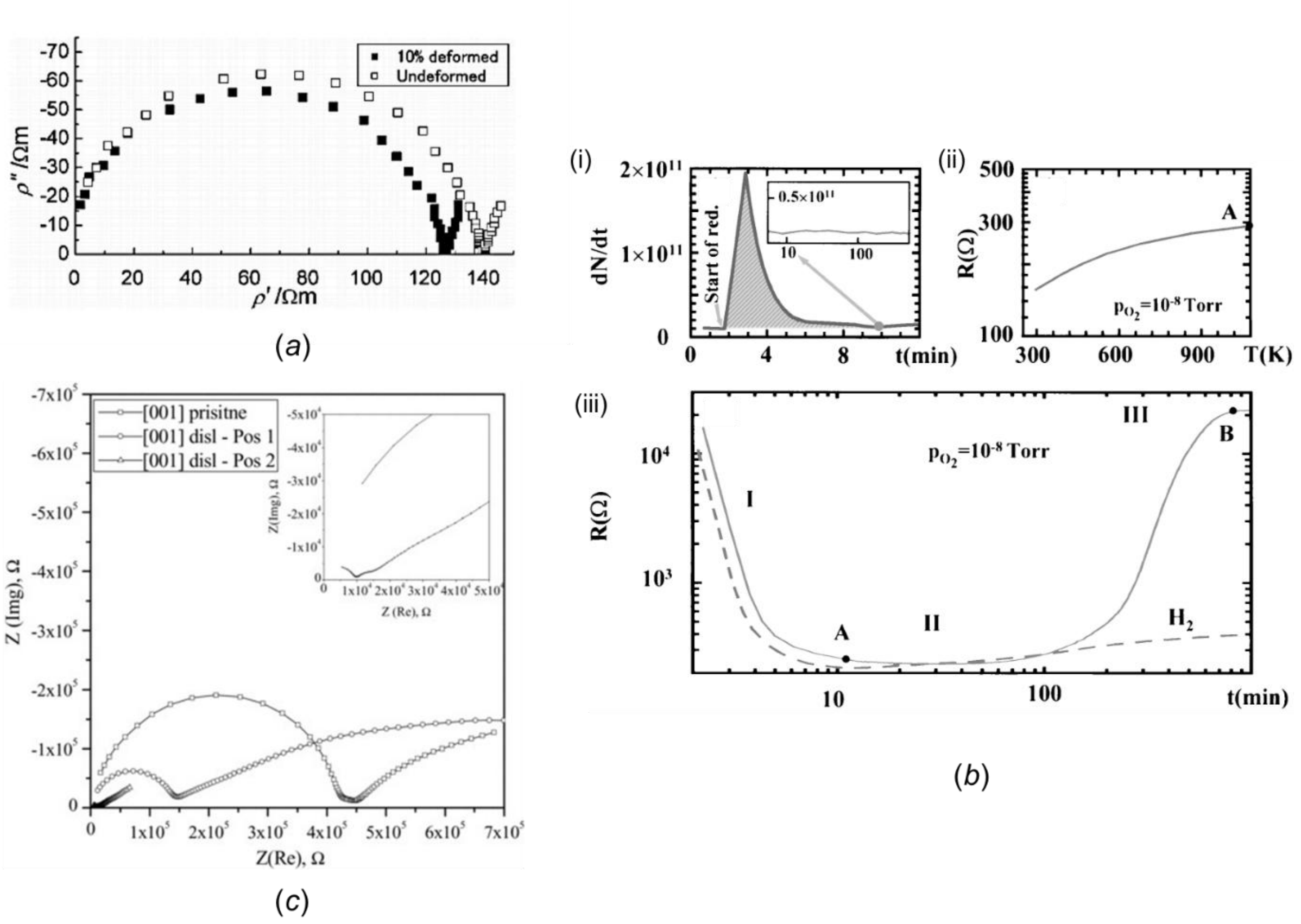

**Fig. 8** Influence of dislocation on resistance of single crystal. (a) Complex impedance plots of YSZ single crystals at 400 °C. Filled squares represent 10% deformed samples. Open squares represent

undeformed samples. The current direction is parallel to the $[\bar{1}\bar{1}1]$ direction [125]. (b) Equilibration kinetics for thermally reduced $SrTiO_3$ under vacuum conditions ( $p_{o_2}$ =1 × $10^{-8}$ Torr, T=800°C) monitored by the oxygen lost from the sample by effusion measurements (i) and independent measurements of the electrical resistivity (ii). Cooling samples after heat treatment at point A reveals metallic behavior down to room temperatures (iii). The sample reduced under $H_2$-enriched atmosphere (dashed line) at the same partial pressure of oxygen shows a similar trend, though markedly different in absolute values for extensive reduction [14]. (c) Impedance spectra of $T_{001}$ sample with and without dislocations at 823 K at 1 bar oxygen. Sample with dislocations show electrode polarization [37].

It has been discovered that dislocations can enhance the conductivity of many ceramic materials such as NiO [148], α-$Al_2O_3$ [149, 150], BZO [151], BZY [152], MgO [153], and $La_{0.8}Sr_{0.2}MnO_3$ [154]. However, the mechanisms behind this phenomenon vary no unanimous, explanations can be categorized as follows： pipe diffusion， strain fields and space charge [155, 156]. The regulation of ceramic conductivity by the dislocation transport mechanism is a promising way beyond the conventional limits of equilibrium bulk doping, and the phenomenon of the structural transport mechanism also has great potential for application at all relevant scales. Therefore, further experiments and simulations are needed to explore the transport properties of ceramic materials in more detail.

### *3.2 Ferroelectricity*

In contrast to alternative materials, ceramics demonstrate a notable superiority in their diverse electromechanical applications, prominently characterized by their ferroelectricity, which enable ceramic materials to be widely used in fields such as sound and vibration sensors, medical ultrasound imaging, precision instruments, electronic devices, and memory devices [157]. Ferroelectricity arises from the polarization behavior of materials, wherein, even in the absence of an external electric field, the material exhibits two or more stable or metastable states with distinct nonzero polarizations. These polarization states are formed through the internal microstructure and atomic arrangement of the material. Ferroelectric materials typically consist of numerous domains, which are microscopic regions with uniform polarization directions. The boundaries between these domains, termed domain walls, determine the overall polarization properties and performance of ferroelectric materials. If the spontaneous polarization directions of two electric domains are perpendicular to each other, they are termed as 90° domain walls. Additionally, there are also 180° domain walls, and so forth, as shown in Fig. 9. Under the influence of an external electric field, the movement of domain walls and rearrangement of domains can result in the reversal of the material's polarization direction, thus achieving ferroelectric property reversal [158]. Therefore, by controlling the size, shape, and distribution of domains, it is possible to alter the dielectric, piezoelectric, and ferroelectric properties of materials, which is of paramount importance in the design and application of ferroelectric materials. Efforts have been continuously made to manipulate the ferroelectric properties of materials to better utilize these advantages.

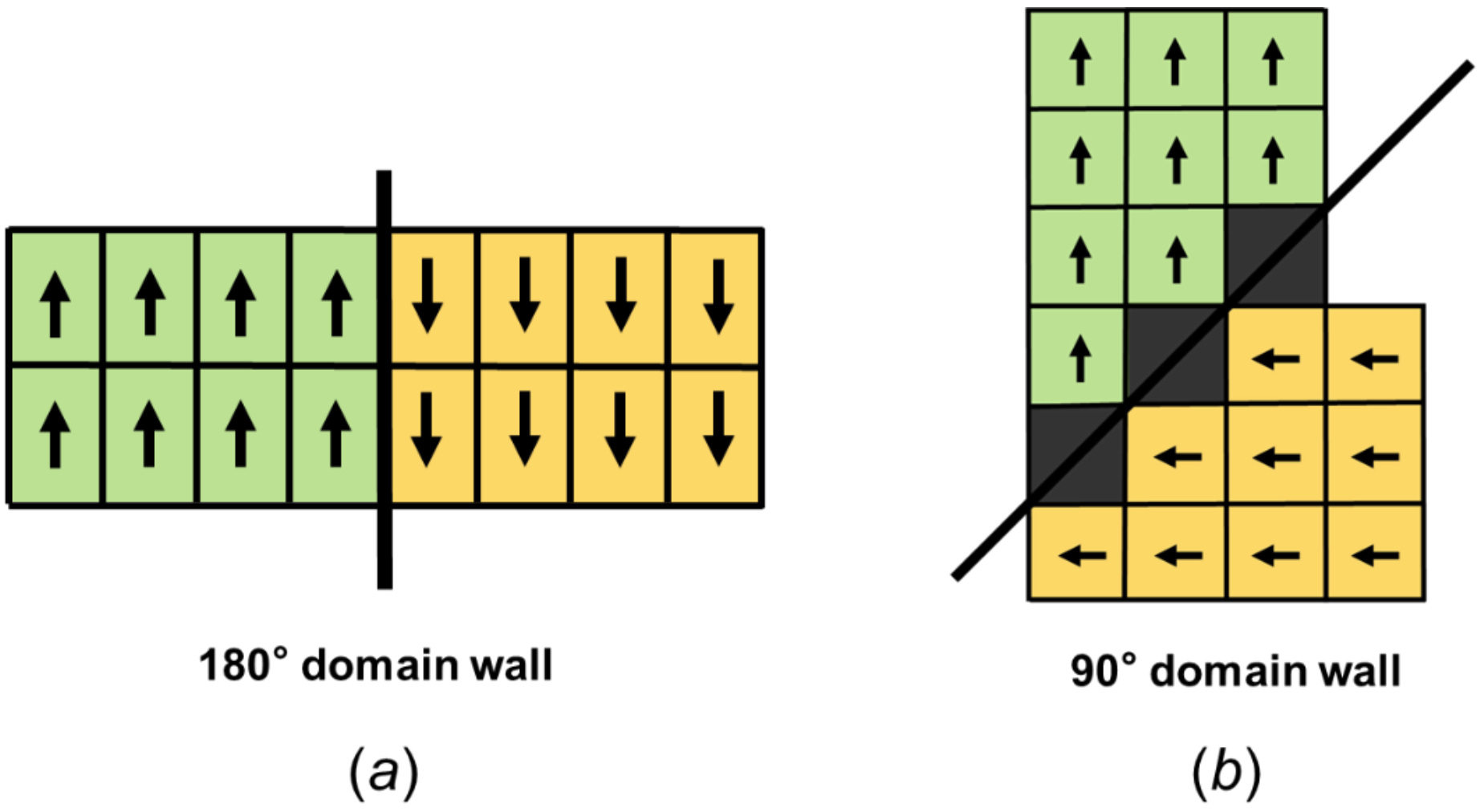

**Fig. 9** Domain walls at different angles. (a) 180-degree domain wall, (b) 90-degree domain wall.

Due to their strong lattice-charge coupling effects, dislocations play a crucial role in determining the piezoelectric and ferroelectric properties. Recent studies have found that dislocations influence ferroelectric properties by altering aspects such as the material's electronic structure, polarization direction, and intensity, thereby providing new avenues for controlling material performance. On one hand, dislocations serve as pinning centers within ferroelectric materials, anchoring localized regions with fixed polarity at domain walls, thereby exerting detrimental effects on the material's polarization [159-161]. Additionally, Jia et al. found that even when dislocations reside outside the ferroelectric material, they can adversely affect the properties of the ferroelectric layers [162]. However, some researchers have observed that during phase transitions, dislocations act as preferential nucleation sites for structural domain formation, thus exerting a positive influence on domain nucleation. Furthermore, they argue that the impact of dislocations on ferroelectric material phase transitions, whether positive or negative, depends on the inherent properties of the material itself [163]. Li et al., through simulation techniques, discovered enhanced local polarization induced by strain around individual dislocations [41]. Shimada et al., via first-principles calculations, demonstrated that due to the inherent local nonstoichiometric of the core structure, Ti-rich $PbTiO_3$ dislocations exhibit magnetism [164]. They also found that by combining the ferroelectric properties of dislocations with those of the matrix, it is possible to create atomically scaled magneto-electric multiferroic materials. Rödel's team has explored various methods to manipulate the ferroelectric coefficients of ceramic materials through dislocation engineering. For instance, they conducted tests on polycrystalline $BaTiO_3$ using high-temperature creep techniques. They found that with a plastic deformation of 1.29%, the Curie temperature increased by 5°C, while the electromechanical strain decreased by 30% [165]. They also creatively combined topological one-dimensional (1D) dislocations and topological two-dimensional (2D) defects domain walls using the method of mechanical dislocation imprint to produce extraordinary and stable $BaTiO_3$ ceramics with anisotropic high dielectric, piezoelectric, and ferroelectric properties [166, 167], as shown in Fig. 10. All these findings offer various insights into manipulating material properties through defect engineering.

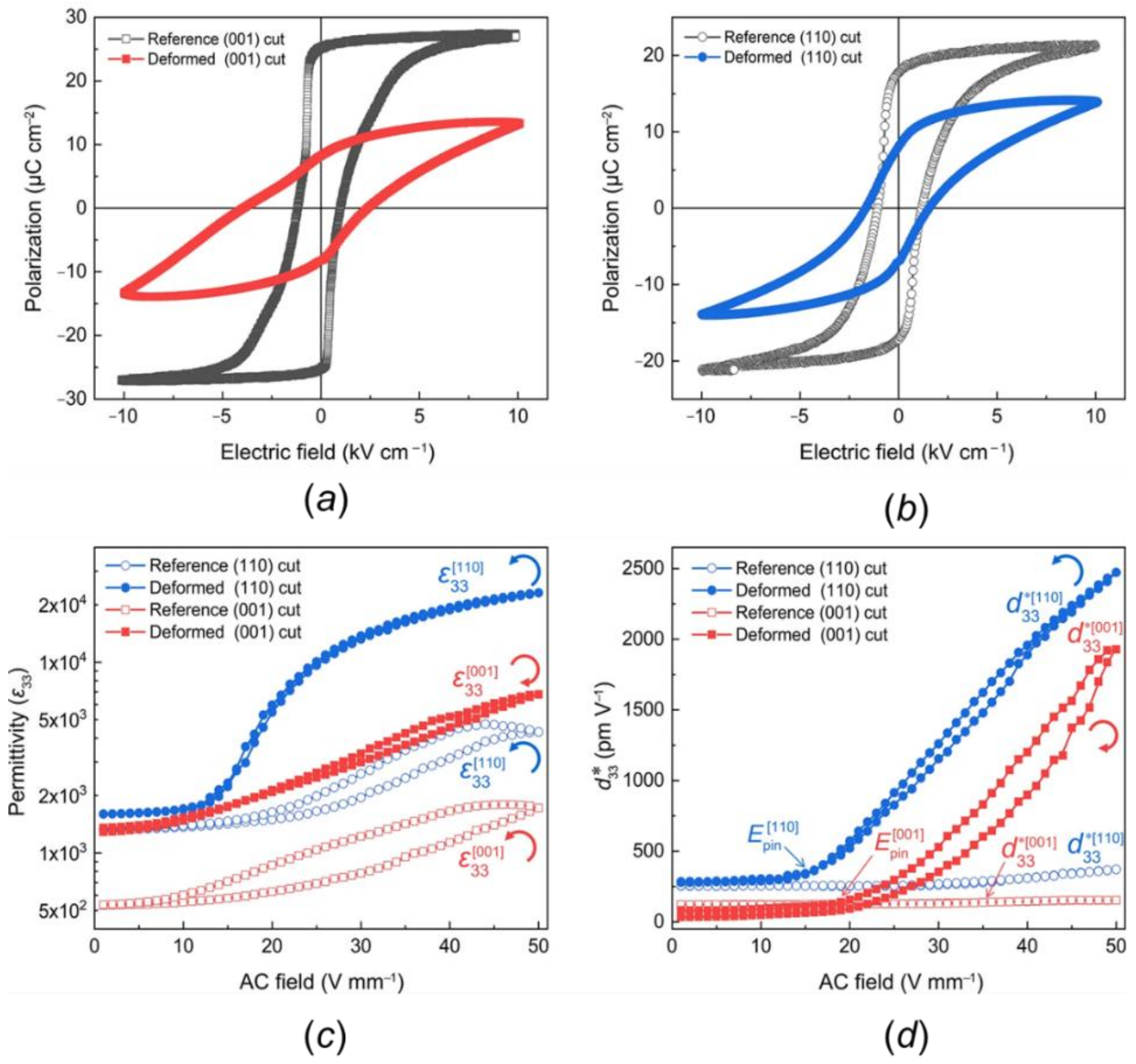

**Fig. 10**. Influence of dislocations on electrical properties. (a, b) Polarization hysteresis loops of reference and deformed (001)- and (110)-cut samples quantified at room temperature with a frequency of 1 Hz. (c) Dielectric permittivity, $\varepsilon_{33}$, and (d) corresponding converse piezoelectric coefficient, $d_{33}^*$, as a function of the amplitude of AC field for reference and deformed (001)- and (110)-cut samples measured at 1 kHz. The inset arrows indicate that the $\varepsilon_{33}$ and $d_{33}^*$ exhibit clockwise (red semicircle arrow) and counter-clockwise hysteresis (blue semicircle arrow) during the AC field cycle. The pinning electric field is defined as the point where $d_{33}^*$ starts to increase dramatically [167].

It is also worthy noticing that one current research hotspot in ferroelectric materials is their application in thin films. With the emergence of micro- and nanotechnologies, various Microelectromechanical Systems (MEMS) and Nanoelectromechanical Systems (NEMS) are focusing on developing materials with thinner thicknesses. However, the influence of dislocations on the performance of small-scale materials is more pronounced, as the relative size of defects increases with decreasing material dimensions. For thin ferroelectric films, a significant issue arises: when the material thickness becomes too low, the ferroelectric properties disappear, greatly hindering the advancement of various technologies [168-170]. Through experiments on (001)-oriented $Pb(Zr_{0.52}Ti_{0.48})O_3$ (PZT) thin film materials, researchers have discovered that misfit dislocations can induce polarization instability in thin ferroelectric materials at the nanoscale [171]. This is because they hinder the formation of stable polarization due to strain fields and lattice deviations within nanoscale islands. Misfit dislocations can accommodate internal strains and spontaneous lattice deformations, leading to the loss of ferroelectricity in nanostructured ferroelectric materials. This indicates that misfit engineering is indispensable for obtaining nanostructured ferroelectric materials with stable polarization.

### *3.3 Thermal conductivity*

Thermal conductivity refers to the inherent ability of a material to transfer or conduct heat, and it plays a crucial role in various engineering and scientific fields. It is typically expressed as the amount of heat (W/ (m·K)) passing through a unit thickness of the material per unit area under a unit temperature gradient. The thermal conductivity $\lambda$ of a homogeneous material can be calculated using

the following formula

$$Q = \frac{\Delta TA}{L}\lambda \quad (4)$$

where $Q$ is the heat flux passing through the conductor, $\Delta T$ is the temperature difference across the conductor, $A$ is the cross-sectional area, and $L$ is the length of the conductor.

In solids, heat conduction primarily occurs through collisions and vibrations of atoms or molecules, facilitated by two main mechanisms: electron conduction and lattice conduction. Electron conduction is predominant in metals, where free electrons move within the crystalline structure, forming an electron gas. Conversely, in ceramics, thermal conduction primarily relies on lattice conduction, where heat is transmitted through the elastic vibrations of the lattice, propagating in the form of phonons (lattice vibrational quanta), resulting in relatively lower thermal conductivity. Ceramic materials offer significant advantages in certain engineering and application fields. For instance, in aerospace applications, lightweight ceramic materials with high temperature resistance and corrosion resistance are required. Additionally, in electronic devices and high-voltage applications, ceramics with electrical insulation properties are essential to prevent potential electronic failures.

Thermoelectric technology finds increasingly widespread applications in various fields such as energy recovery, portable electronic devices, and wireless sensors [172]. Therefore, there is a need to strike a balance between enhancing electrical conductivity while reducing thermal conductivity. Additionally, since thermal conductivity is relatively independent of electrical conductivity [173], reducing thermal conductivity has become a key technological strategy to meet the market demand for more efficient thermoelectric technology. It has long been observed that introducing either ordered or disordered dislocations can effectively reduce the phonon mean free path (MFP), thereby lowering thermal conductivity and providing better control over the thermoelectric coefficient. This phenomenon has been confirmed through years of continuous experimentation and simulation for both alloy [23, 174] and ceramic [20, 175-178]. For instance, Yokota et al. utilized $Yb_2O_3$ and $ZrO_2$ as sintering additives to produce two types of β-$Si_3N_4$ with a 40 (W/ (m·K)) difference in thermal conductivity by varying the sintering conditions [179]. It was observed that in the ceramic with lower thermal conductivity, dislocations were prominently visible, whereas conversely, dislocations were absent in the higher thermal conductivity material. This observation confirms that dislocations indeed lower the thermal conductivity of materials. Similarly, Johanning et al. introduced a large number of dislocations into $SrTiO_3$ using plastic deformation [20]. Through testing, they found that both single crystals and polycrystals of $SrTiO_3$ exhibit a significant decrease in thermal conductivity as the dislocation density increases. Other scholars from Japan, utilizing molecular dynamics simulations, discovered that in MgO, the contributions of edge and screw dislocations to thermal conductivity differ significantly [177]. The thermal conductivity is nearly independent of the length of edge dislocations (Fig. 11a), whereas with increasing length of screw dislocations, a significant decrease in thermal conductivity was observed, markedly lower than that under pure edge dislocations (Fig. 11b). Khafizov et al. introduced faulted dislocation loops with diameters of a few nanometers into polycrystalline $CeO_2$ samples by irradiating them with 1.6 MeV protons at 700°C [178]. The thermal conductivity of the damaged layer was measured using modulated thermoreflectance. The data indicate that the reduction in thermal conductivity is primarily attributed to the dislocation loops, while the influence of point defects and vacancies is minimal, as depicted in Fig. 11(c).

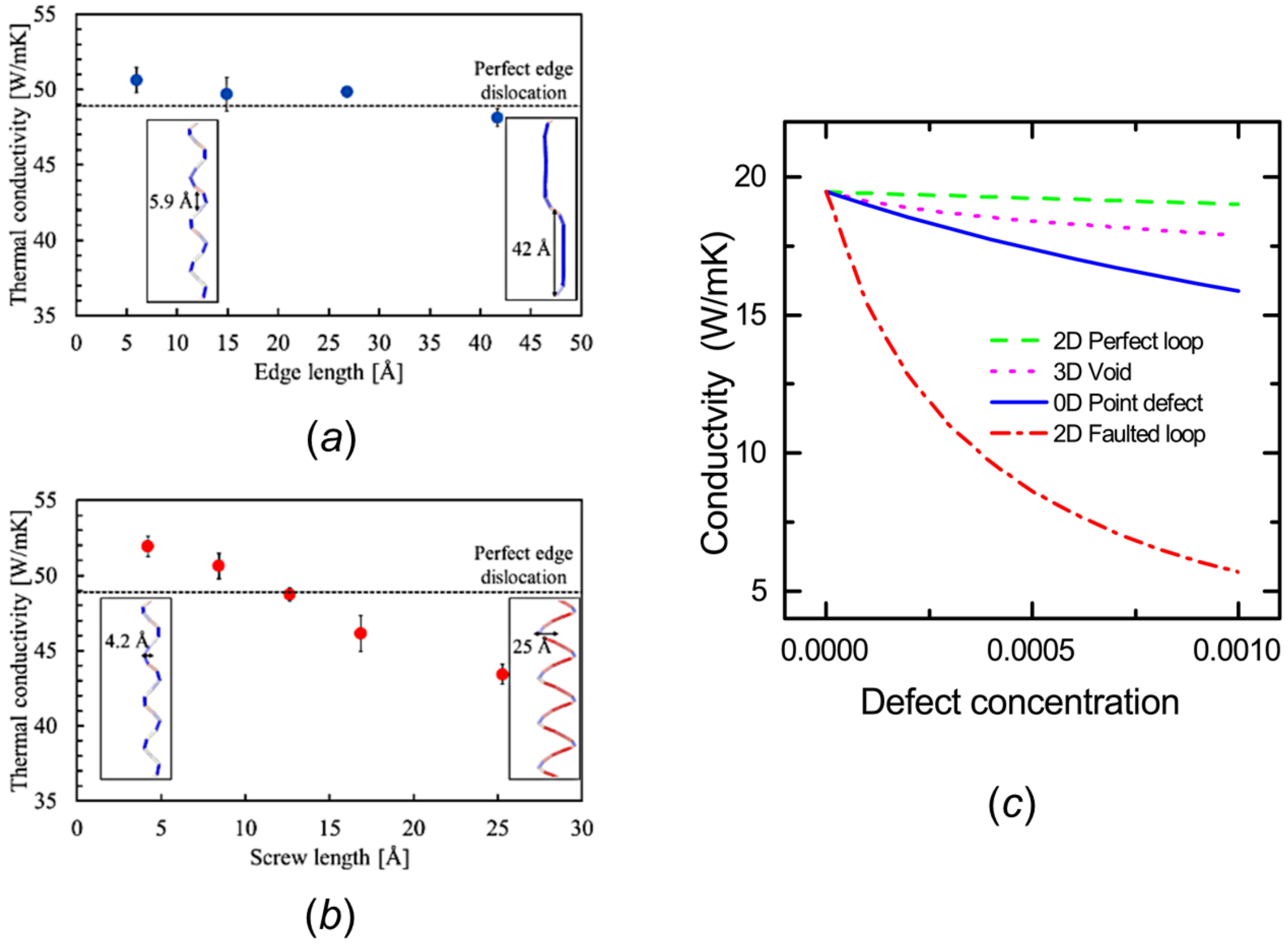

**Fig. 11** Thermal conductivity as a function of (a) edge length and (b) screw length. The screw and edge length are fixed to about 8 and 6Å in (a) and (b), respectively. The thermal conductivity with perfect edge dislocation is also shown as a dashed line. (c) Impact of irradiation induced defects on conductivity reduction [177, 178].

## 3.4 Superconductivity

Superconductivity, a phenomenon initially observed in the early 20th century, describes the remarkable property of certain materials to exhibit zero electrical resistance and expel magnetic fields from their interior upon cooling to a critical temperature [180, 181]. The theoretical framework underpinning superconductivity is elucidated by the Bardeen-Cooper-Schrieffer (BCS) theory, which posits the formation of electron pairs within superconductors, allowing them to move without hindrance and manifesting the zero-resistance characteristic.

The applications of superconductivity span across various domains such as MRI, superconducting quantum interference devices, superconducting microwave devices, and superconducting quantum computing. However, the practical utilization of superconductivity is impeded by the typically low transition temperatures, motivating ongoing efforts to discover materials with lower transition temperatures or develop methods to enhance existing ones. High-temperature superconductivity, characterized by critical temperatures surpassing the boiling point of liquid nitrogen (77K), predominantly manifests in ceramic forms. Nevertheless, the understanding of the mechanism behind high-temperature superconductivity remains incomplete within the current scope of research, with many phenomena defying conventional explanations offered by the BCS theory.

Recently, dislocations have been recognized as a means to manipulate the superconducting properties of materials [18, 182, 183]. For instance, in $YBa_2Cu_3O_{7-\delta}$, Dam et al. found that dislocations can provide strong pinning centers responsible for high critical currents, effectively restricting the flow of superconducting currents and increasing the stability of the superconductor [183]. The superconductivity of $SrTiO_3$ was discovered as early as 1964 [184], yet its underlying principles remain largely unexplored. The superconductivity of strontium titanate cannot be explained by the electron–phonon pairing mechanism because it occurs at extremely low charge-carrier densities [185, 186]. Surprisingly, it has been observed that superconductivity in strontium titanate can be enhanced through plastic deformation, raising the superconducting transition temperature by up to two orders of magnitude [18]. This phenomenon has been experimentally confirmed through diffuse neutron and X-ray scattering to originate from dislocations generated during plastic deformation.

## 4 Conclusions and perspectives

The potential of dislocation manipulation in ceramic performance might have been significantly underestimated in the past. In recent years, with the development of science technology and theory, mutually corroborative experiments and simulation results have brought this field into focus. The field of dislocations in ceramics is highly promising, with a broad scope and promising results across multiple areas, including enhanced plasticity and fracture toughness, improved electrical conductivity, and ferroelectric properties, among others. Nevertheless, research in most of these areas is still in its infancy, with considerable controversies and open questions concerning the underlying mechanisms.

The most direct and significant impact of dislocations on ceramic performance is reflected in their mechanical behavior. Dislocations glide, multiply, and climb in ceramics, contributing to plasticity and regulating fracture toughness and damage tolerance. Various methods have been explored to enhance the mechanical performance of ceramics, including adjusting temperature, modifying the microstructure of ceramic constituents, and changing dopants and their concentrations. Research in the field of mechanical behavior is well-established. However, another research direction focused on dislocation-tuned functional properties has attracted considerable attention but remains highly controversial. Different ceramic materials, synthesis methods, and modes of dislocation generation lead to varied outcomes, with the underlying mechanisms behind these phenomena still under debate. As a result, the study of transport properties holds significant potential for further exploration. From an application perspective, electromechanical properties, particularly in ferroelectrics, represent a crucial area of research for the utilization of functional ceramics. While dislocations are often thought to negatively affect ferroelectric behavior, it is increasingly recognized that careful manipulation of dislocations can also yield favorable results in ferroelectric properties. Finally, we discuss thermoelectric and superconducting properties, where research is still ongoing but has shown promising results. Dislocations have been found to enhance ceramic performance in these areas.

Given the complexity and diversity of the field of dislocations in ceramics, we expect that by summarizing research results and underlying control mechanisms from different directions, it can facilitate interdisciplinary research efforts to advance this field.

**Acknowledgement:**

X. Liang are grateful for the support by the National Natural Science Foundation of China (No. 12122209, 12072251, and 12472157), Project B18040, and the Young Talent Plan of Xi'an Jiaotong University. X. Fang acknowledges funding by the European Research Council (ERC) under Grant No. 101076167 (Project MECERDIS). Views and opinions expressed are, however, those of the authors only and do not necessarily reflect those of the European Union or the European Research Council. Neither the European Union nor the granting authority can be held responsible for them.